\documentclass{article}
\usepackage[utf8]{inputenc}
\usepackage{xcolor}
\usepackage[a4paper,width=150mm,top=25mm,bottom=25mm]{geometry}
\usepackage{hyperref}
\usepackage{graphicx}
\usepackage{amssymb}
\usepackage{amsmath}

\definecolor{newcolor}{rgb}{.8,.349,.1}
\definecolor{shadecolor}{cmyk}{0,0,0,.1}

\usepackage{subfig}
\usepackage[symbol]{footmisc}
\usepackage{authblk}
\usepackage{multirow}

\newif\ifdeleted

\DeclareMathOperator*{\argmin}{argmin}
\date{}

\title{Identifying Signatures of Image Phenotypes to Track Treatment Response in Liver Disease}

\author[1, 2]{Matthias Perkonigg}
\author[1]{Nina Bastati}
\author[1]{Ahmed Ba-Ssalamah}
\author[3]{Peter Mesenbrink}
\author[3]{Alexander Goehler}
\author[4]{Miljen Martic}
\author[3]{Xiaofei Zhou\footnote{Present address: Alnylam Pharmaceuticals, Cambridge, MA, USA}}
\author[5]{Michael Trauner}
\author[1,6]{Georg Langs\footnote{corresponding author: Medical University of Vienna, Waehringer Guertel 18–20, 1090 Vienna, Austria. georg.langs@meduniwien.ac.at}}

\affil[1]{Department of Biomedical Imaging and Image-guided Therapy, Medical University of Vienna, Austria}
\affil[2]{Institute of Clinical Epidemiology, Public Health, Health Economics, Medical Statistics and Informatics, Medical University of Innsbruck, Austria}
\affil[3]{Novartis Pharmaceuticals Corporation, East Hanover, NJ, USA}
\affil[4]{Novartis Pharma AG, Basel, Switzerland}
\affil[5]{Division of Gastroenterology and Hepatology, Department of Medicine III, Medical University of Vienna, Austria}
\affil[6]{Comprehensive Center for Artificial Intelligence in Medicine, Medical University of Vienna, Austria}

\begin{document}
\maketitle

\begin{abstract}
Quantifiable image patterns associated with disease progression and treatment response are critical tools for guiding individual treatment, and for developing novel therapies. Here, we show that unsupervised machine learning can identify a pattern vocabulary of liver tissue in magnetic resonance images that quantifies treatment response in diffuse liver disease. Deep clustering networks simultaneously encode and cluster patches of medical images into a low-dimensional latent space to establish a tissue vocabulary. The resulting tissue types capture differential tissue change and its location in the liver associated with treatment response. We demonstrate the utility of the vocabulary on a randomized controlled trial cohort of non-alcoholic steatohepatitis patients. First, we use the vocabulary to compare longitudinal liver change in a placebo and a treatment cohort. Results show that the method identifies specific liver tissue change pathways associated with treatment, and enables a better separation between treatment groups than established non-imaging measures. Moreover, we show that the vocabulary can predict biopsy derived features from non-invasive imaging data. We validate the method on a separate replication cohort to demonstrate the applicability of the proposed method.
\end{abstract}


\newpage
\section{Introduction}
\label{sec:introduction}

Non-invasive imaging such as magnetic resonance imaging (MRI) can obtain quantitative local features, as well as their spatial distribution in organs associated with disease. This is useful for diagnosis, the assessment of disease progression and treatment response, or the prediction of future patient trajectories to guide treatment decisions. Identifying such features - \textit{markers} - and establishing their robust relationship with disease staging and treatment response is challenging. Compared to biopsy-derived markers, they reduce patient risk, enable repeated measurements over the course of treatment, and capture possibly heterogeneous distributions of disease patterns. 

Here, we propose a novel approach for marker discovery in the form of a visual pattern vocabulary. We demonstrate that it enables sensitive quantitative assessment of treatment response and is able to capture heterogeneous localized tissue change over time. 

%

\subsection{The need for non-invasive markers}

Non-alcoholic steatohepatitis (NASH) is a progressive liver disease and a severe form of nonalcoholic fatty liver disease (NAFLD) \cite{Drescher2019CurrentNASH}. Despite different non-invasive imaging features and clinical factors related to NASH, the gold standard for accurate staging remains invasive biopsy \cite{Wong2019DiagnosticSteatohepatitis}. Aside from being invasive, biopsy is prone to sampling errors \cite{Ratziu2005SamplingDisease} and inter- and intra-reader variability leading to misdiagnosis~\cite{Vuppalanchi2009EffectsDisease}. 

Therefore, it is an example for the need of non-invasive repeated quantitative assessment~\cite{Drescher2019CurrentNASH}, and the expansion of our vocabulary to capture complex changes of multiple tissue characteristics during disease progression and treatment response. NASH is characterized by three important factors used for diagnosis: the accumulation of fat within the liver (steatosis), hepatocellular ballooning, and lobular inflammation. Staging liver fibrosis is important as a marker of disease progression~\cite{Wong2019DiagnosticSteatohepatitis}. 

Current treatment of NASH involves encouraging patients to lifestyle and dietary changes, but a pharmacological treatment is needed due to inconsistent long-term success \cite{Drescher2019CurrentNASH}. During routine treatment, and clinical trials, means for monitoring and reliable endpoints to evaluate the effectiveness of treatment are highly relevant. Established invasive methods relying on biopsy are unsuitable. Current non-invasive markers to measure fat accumulation or fibrosis in NAFLD patients are weak in differentiating between disease stages and are not suitable to track disease progress~\cite{Wai-SunWong2018NoninvasivePromise}.

MRI is non-invasive, captures the entire liver, and is not prone to sampling errors like biopsy. It is suitable for follow-up measurements at corresponding locations to capture subtle tissue changes and could enable more precise monitoring of progression under treatment.

\subsection{Contribution}

Here, we propose an approach to capture heterogeneous changes visible in liver MRI associated with treatment response and disease stages of NASH. We perform unsupervised deep convolutional clustering on image patches extracted from multi-parametric liver MRI, to establish a vocabulary of liver tissue properties and their appearance in multi-parametric MRI. Simultaneous dimensionality reduction and clustering identifies frequently occurring and discernible appearance patterns. 
From those patterns we build pattern profiles and demonstrate that first, the approach is capable to discover pattern profiles that enable the quantification of treatment response with higher accuracy than established markers. Second, we track changes of patterns at corresponding locations over time to identify differential tissue change pathways associated with treatment response. Finally, we learn a mapping from these pattern profiles to markers typically evaluated in histo-pathology from biopsy to demonstrate their ability for non-invasive quantitative assessment. We validate this experiment on a separate replication cohort.

\section{Related work}

\paragraph{Machine learning in NASH} Machine learning methods for assessing NASH markers in histo-pathology \cite{Heinemann2019DeepModels, Arjmand2020TrainingSamples} or on electronic health records (EHR)~\cite{Fialoke2018ApplicationPatients} have shown promising results. However, histo-pathology is invasive and thus not suitable for monitoring during the treatment of the disease. Existing work includes the  distinction between NASH and non-NASH patients without predicting the patients diseases status and progress~\cite{Fialoke2018ApplicationPatients}. Deep learning has been shown to predict the fibrosis grade in contrast enhanced computer tomography with high accuracy~\cite{Choi2018DevelopmentLiver}. In \cite{Bastati2023} it was shown that features extracted with a deep learning method can differentiate between simple steatosis and NASH.

\paragraph{Unsupervised clustering} Several unsupervised clustering methods for natural images exist. They can be broadly divided into auto-encoder (AE) based approaches \cite{Yang2017TowardsClustering, Xie2016UnsupervisedAnalysis, Li2017DiscriminativelyAuto-Encoders, Dizaji2017DeepMinimization}, generative model based methods \cite{Jiang2017VariationalClustering} and direct cluster optimization techniques \cite{Yang, Chang2017DeepClustering, Hu2017LearningTraining}. AE based methods augment the standard AE reconstruction loss with a second loss term to facilitate clustering in the latent space. Generative model based approaches build on variational auto encoders or generative adversarial networks and also add a clustering loss. In contrast, direct cluster optimization methods rely on a clustering loss only to optimize the feature extracting deep neural network.
We use an AE method building on Deep Clustering Networks (DCN) \cite{Yang2017TowardsClustering} to jointly learn and cluster latent space features across different sequences in multi-parametric MRI.
Deep clustering in conjunction with medical imaging has been used for unsupervised \cite{Moriya2018UnsupervisedLearning} and semi-supervised \cite{Enguehard2019Semi-supervisedSegmentation} segmentation of lung tissue and brain regions, respectively. Moriya et al. \cite{Moriya2018UnsupervisedLearning} used JULE, a direct cluster optimization approach \cite{Yang} to segment micro CT lung images into invasive carcinoma, non-invasive carcinoma and normal tissue. Different from our work, a cluster embedding is learned for each case separately and clusters are not shared over the whole population.
The semi-supervised approach \cite{Enguehard2019Semi-supervisedSegmentation} uses a clustering layer based on t-distributed stochastic neighborhood embedding (TSNE) on top of a feature extracting convolutional neural network (CNN) to reduce the amount of data needed for training a segmentation algorithm. They showed that with only a fraction of labeled training data that their method is able to segment white matter, gray matter and cerebrospinal fluid from MRI T1w and T2w images. These segmentation approaches aim to reduce the amount of labeling required by the radiologist. In contrast, we use deep clustering to identify predictive patterns, possibly be unknown to medical experts.
Li et al. \cite{Li2018AlzheimersNetworks} use K-Means clustering on image patches of brain MRI to extract features for Alzheimer's disease classification.
Wang et al. \cite{Wang2017UnsupervisedRecognition} showed a deep clustering approach for medical image categorization. They cluster features extracted by a CNN on images to divide images into clusters of similar anatomy or pathology. The results in this paper have been developed as part of the author's thesis \cite{PerkoniggThesis}.

\section{Method}
Our method is composed of an unsupervised clustering step to learn an appearance vocabulary and a subsequent supervised step to discover disease and progression markers within the learned vocabulary.

\begin{figure}[ht!]
\centering
\includegraphics[width=0.8\textwidth]{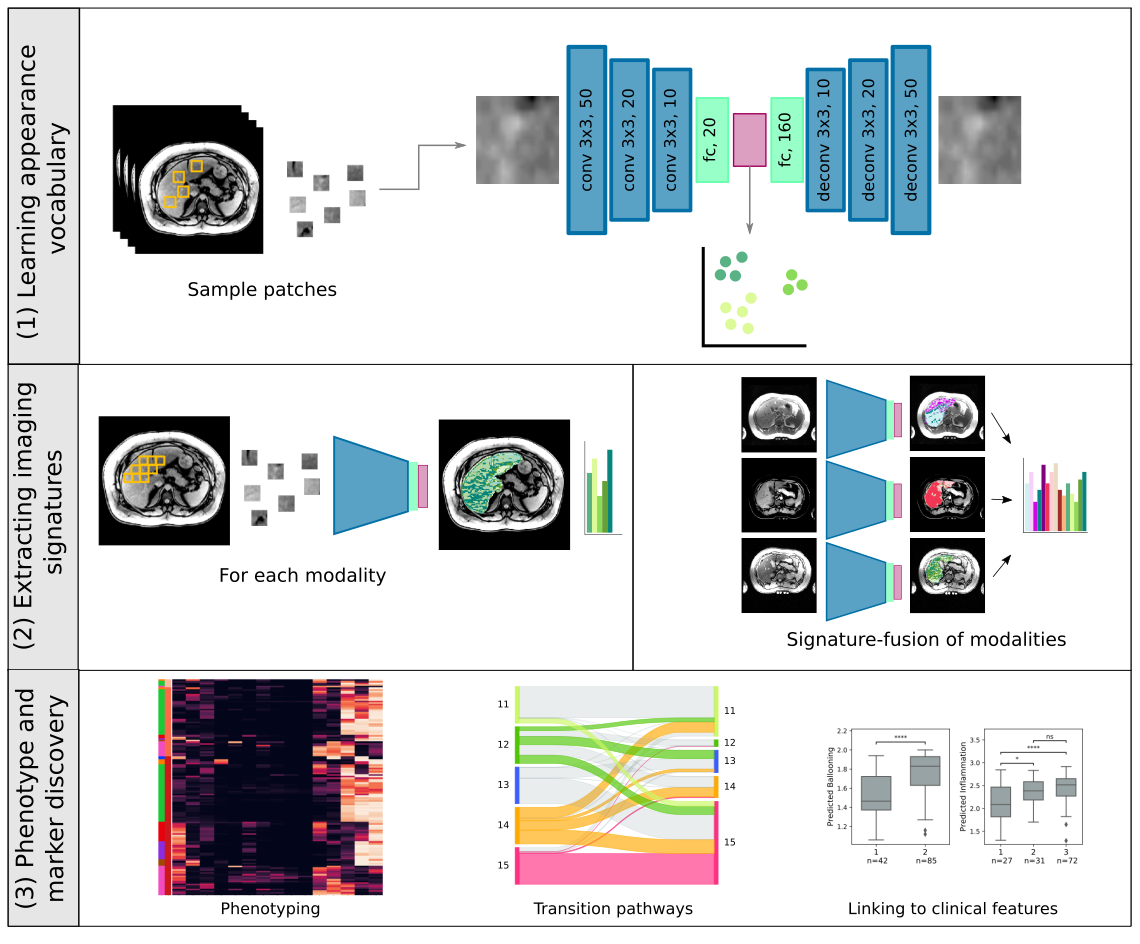}
\caption{Overview of the proposed method. To train the DCN, image patches are sampled randomly from the liver of all patients and the DCN is trained. The trained encoder of the network is applied to each patient by extracting and encoding a patch to a cluster at each position within the liver, the relative frequency of clusters is used as \textit{image signature}. The set of \textit{signatures} of all patients is used for multiple exploratory tasks: 
Phenotyping, progression marker discovery and biopsy prediction.}
\label{fig:method}
\end{figure}

\subsection{Learning an appearance vocabulary from multi-parametric liver MRI}

We learn a vocabulary of liver tissue and corresponding appearance acquired by a set of sequences in multi-parametric MRI (Fig.\,\ref{fig:method}). We (1) sample image patches to compose a training data set, then (2) deep clustering networks learn a representation of the image patches~\cite{Yang2017TowardsClustering}. For new imaging data (3) the trained models extract \textit{image signatures} describing the composition of the imaged liver. 

\subsubsection{Extracting patches from volumes}
Given a set of $N$ volumes $\mathcal{V} = \{\mathbf{v}_1, \dots, \mathbf{v}_N\}$ with corresponding binary segmentation masks of the liver $\mathcal{M} = \{\mathbf{m}_1, \dots, \mathbf{m}_N\}$, a set of $M$ image patches $\mathcal{P} = \{\mathbf{p}_1, \dots, \mathbf{p}_M\}$ is extracted as a training set. For each volume $\mathbf{v}_i \in \mathcal{V}$, $\lfloor \frac{M}{N} \rfloor$ patches are extracted at random positions within the masked liver $\mathbf{m}_i$. The patches are drawn in 2D along the axial slice with a patch size of $s$. A 2D patch is extracted at the same position in every echo, if the MRI sequence evaluation has multiple echos. Afterwards, those echo patches are concatenated along the channel axis to form a multi-channel 2D patch.
The set of extracted patches $\mathcal{P}$ is used to train the clustering networks.

\subsubsection{Deep Clustering Networks (DCN) for medical imaging patches}
\label{sec:method_DCN}
In this section, we will first review the approach by \cite{Yang2017TowardsClustering} which has been developed for  different benchmark data sets using auto encoders without convolutions. We expand the idea to be used on patches of medical images. 
\\
DCN is based on auto encoders using the bottleneck layer to identify clusters of similar inputs. The input is a patch $p \in \mathcal{P}$ encoded by the encoder $\mathbf{f}$ to a latent representation $z=\mathbf{f}(p)$. Then the decoder $\mathbf{g}$ attempts to reconstruct the input patch by decoding $z$ to the output $y=\mathbf{g}(z)=\mathbf{g}(\mathbf{f}(p))$. By exploiting correlations and redundancies in $p$, $z$ is a compressed representation of $p$ and encodes distinct features. Classical auto encoders are trained by minimizing a reconstruction loss to learn how to reconstruct $p$ from $z$. The DCN method also uses a reconstruction loss, in this work we train the network using the mean squared error:
\begin{equation}
\label{eq:reconloss}
\mathcal{L}_{recon}= \frac{1}{N}\sum_{n=1}^N (p_i-\mathbf{g}(\mathbf{f}(p_i)))^2
\end{equation}
where $\theta$ and $\phi$ are the parameters of $\mathbf{f}$ and $\mathbf{g}$ respectively.

To enforce the clustering in the latent space DCN uses a second loss term measuring the \textit{cluster-friendliness}, which is derived from the k-means algorithm \cite{Lloyd1982}. This loss aims to minimize the euclidean distance between the current cluster centroids $C$ and the training samples assigned to that cluster:
\begin{equation}
\label{eq:kmeans}
\mathcal{L}_{cluster}=\Vert \mathbf{f}(p_i) - \mathit{C}m_i\Vert_2^2
\end{equation}
where $m_i$ is a one-hot encoded vector denoting the cluster membership for each patch $p_i$.
The full loss function can be written as
\begin{equation}
\label{eq:loss}
\mathcal{L} = \mathcal{L}_{recon} + \lambda \mathcal{L}_{cluster}
\end{equation}
with $\lambda$ being a hyperparameter that regulates the trade-off between reconstruction quality and \textit{cluster-friendliness} (\cite{Yang2017TowardsClustering}).

For a fixed number of epochs a pre-training with only the reconstruction loss according to Eq. \ref{eq:reconloss} is used. Afterwards the clustering in latent space is initialized by performing k-means assigning each $p_i$ to a cluster.
Next, the network is trained in an alternating gradient descent scheme of updating the network parameters by Eq. \ref{eq:loss}, updating the cluster assignments of the training samples by
\begin{equation}
m_{i, j}= 
\begin{cases}
    1, & \text{if } j = \argmin_{k \in \{1,\dots,K\}} \Vert \mathbf{f}(x_i) - c_k\Vert_2 \\
    0,              & \text{otherwise}
\end{cases}
\end{equation}
and updating the cluster centroids by
\begin{equation}
    c_k = c_k - \frac{1}{n_k}(c_k-\mathbf{f}(x_i))m_{i,k}\textrm{,}
\end{equation}
where $n_k$ is the number of samples currently assigned to $c_k$.
For more details on the training procedure see \cite{Yang2017TowardsClustering}.

\paragraph{Auto-encoder architecture}
The choice of the network architecture is an important step. We use a simple convolutional auto encoder: a 32x32 input is run through three convolutional layers (50, 20 and 10 feature maps respectively) with receptive field of 3x3 and ReLU activation followed by a 2x2 MaxPooling. The features are reduced to a 20 dimensional latent space vector by a fully connected layer. The decoder performs deconvolution by upsampling followed by a convolutional layer in the reversed order. Our network is depicted in Figure \ref{fig:method}.

\subsubsection{Multi-sequence image signatures}
\label{sec:method_fusionstrategies}
After the DCN is fully trained, we use a sliding window approach to parse through the ROI of a volume $v_i \in \mathcal{V}$ by extracting a set patches at each position within the ROI ($\mathcal{P}_i$).
Each of those patches $p_j \in \mathcal{P}_i$ is first encoded ($\mathbf{f}(p_j)$) to the latent space and afterwards assigned to the nearest cluster by minimizing

\begin{equation}
 c(p_j) = \argmin_k \Vert \mathbf{f}(p_j) - c_k m_j\Vert_2 | k \in \{1,\dots,K\},
\end{equation}
where $K$ is the number of clusters. After all patches are assigned to a cluster we can either view that map as an overlay of the liver (\textit{cluster map}) or build a compact feature vector to represent the whole image, we call this representation the \textit{signature} ${\bf s}$ of an image. Analogously to the bag-of-visual-words approach \cite{Sivic2003VideoVideos}, this \textit{signature} is a vector where each element $s_i$ is defined as the relative frequency of cluster $i$ within the cluster map

\begin{equation}
{\bf s} = \langle s_1, s_2, \dots, s_K \rangle, \quad  s_i = \frac{\sum_{j=V}^K{c(p_j)==i}}{V},
\end{equation}
%
where $V$ is the number of clusters. We denote the signature for a volume $n$ as ${\bf s}_n$.

To utilize multi-parametric MRI, \textit{\textbf{signature-fusion}} is used to combine signatures from different sequences:  For each sequence $u$ used, a separate \textit{signature} vector ${\bf s}^u$ is calculated. Vectors are concatenated to form the multi-sequence signature-fused signature of a patient visit ${\bf s}_{SF} = \langle {\bf s}^1, {\bf s}^2, \dots, {\bf s}^U \rangle$. In the following we use \textit{SF}, followed by the number of clusters per sequence to refer to such features, e.g. \textit{SF-5-3} are signature-fused features with $K=5$ clusters for each of $U=3$ sequences.

For an alternative merging approach see Section \ref{sec:ablation_study}.

\subsection{Discovering markers of progression and response}

Based on the signatures of livers, marker discovery is performed as supervised training to associate the pattern vocabulary to target variables. 
From the signatures we (1) discover phenotypes in the imaging data of the patient population by clustering in the signature space. In addition, we show how to (2) use the vocabulary to quantify and compare treatment response, and (3) identify pathways of liver tissue transitions as change patterns during treatment. Finally, we (4) predict markers extracted from histo-pathology by supervised learning to confirm that the markers capture clinically meaningful characteristics.

After preliminary experiments, we used random forests \cite{Breiman2001RandomForests} as the supervised machine learning method to achieve the mapping. 
Random forests showed equal performance in terms of accuracy to other models such as support vector machines and enabled the identification of multiple, interpretable predictors.

\subsubsection{Identification of markers capturing treatment response}
To find signature components that are markers for the progression of the disease under treatment or placebo we calculated signatures for longitudinal liver imaging data. The difference of signatures between the initial visit of a patient and a follow-up visit is calculated for each patient. The resulting \textit{difference signatures} ${\bf s}_n^{\bigtriangleup}$ are used to train a random forest to predict the treatment group a patient is assigned to ${\bf s}_n^{\bigtriangleup} \mapsto t_n$ (e.g., $t_n\in \{\textrm{high dose}, \textrm{low dose}, \textrm{placebo}\}$, Section \ref{sec:results_progressmarker}).

\subsubsection{Identification of liver tissue transitions under treatment}
We register follow-up images and corresponding \textit{cluster maps} using 3D rigid registration to identify liver tissue patterns that change under treatment. Tissue transitions from initial visit $t=0$ to follow-up visit $t=1$ are recorded and summarized from the registered \textit{cluster maps} at each position in the liver, resulting in a cluster transition matrix 
\begin{equation}
{\bf M} \in \mathbb{R}^{K\times K}, \quad {\bf M}_{ij} = p\left( c(p_j^{t=0})=i \cap c(p_j^{t=1})=j\right). 
\end{equation}
${\bf M}$ holds transition probability of a tissue changing between initial visit and follow-up. For ${\bf s}_{SF}$ this transition probabilities are calculated for each sequence independently (Section \ref{sec:results_tissuetransitions}).

\subsubsection{Prediction of histo-pathology features}

We employ random forest classification and regression to link histology grades ($h$) to the extracted signatures for an image $n$, ${\bf s}_n \mapsto h_n$. We used binary random forest classifiers to separate patients with low (grade 0, 1) and high grade (\textgreater 1) biopsy values for \textit{inflammation}, \textit{ballooning}, \textit{steatosis}, and \textit{fibrosis} or \textit{NAFLD fibrosis} score respectively. Second, random forest regression was used to evaluate if the signatures captures a progressive trend with increasing biopsy grades.
Linking the extracted signatures to established and clinically relevant grades demonstrates that unsupervised learning generates a meaningful vocabulary (Section \ref{sec:results_prediction}).

\subsubsection{Discovery of phenotypes}
Image phenotypes are groups of patients with similar \textit{image signatures}.
We use hierarchical, agglomerative clustering with average linkage and euclidean distance \cite{Mullner2011ModernAlgorithms} to find similar signatures of patients in the study population $\{ {\bf s}_1, {\bf s}_2, \dots, {\bf s}_N \}$ and assign each patient to a phenotype ${\bf s}_n \mapsto p_n, p_n\in \{1,\dots,P\}$. We form between $P=5$ and $10$ phenotypes to keep the number of phenotypes interpretable. We calculate the Odds ratio (OR) expressed in patients with histo-pathology gradings for each of the discovered phenotypes to evaluate the association between phenotype and known histo-pathology grades
~(Section \ref{sec:results_phenotypes}). 

\section{Experiments and Results}
\label{sec:experiments_results}
\subsection{Data}
We used two independent data sets to evaluate our approach. One data set was collected during a randomized controlled trial (RCT) including NASH patients \cite{Lucas2020TropifexorResults}. To evaluate if the methodology can be applied to different imaging settings, a different data set was extracted from clinical routine data. 

\subsubsection{Randomized Controlled Trial (RCT) data sets}
\label{sec:data_rct}
The RCT data set was collected during FLIGHT-FXR, a phase 2 randomized, double blind, placebo-controlled, 3-part study for several doses of tropifexor in patients with NASH \cite{Lucas2020TropifexorResults}. The study was carried out in multiple centers and MRI was used to monitor hepatic fat fraction (HFF). Patients underwent MRI scanning at the inital visit and at a 12 weeks follow-up. The RCT includes data from 93 MRI scanners. In this work we only included data from scanners of patients who received MRI scanning with three different series: (1) Axial 3D T1-weighted GRE, breath-holding series (depending on scanner manufacturer VIBE, SPGR or T1FFE), in the following abbreviated by \textit{T1w}, (2) Axial 3D Dixon series where fat and water phase images are reconstructed (\textit{dixon}) and (3) a HFF T1-weigthed GRE six echo series, with echo time (TE) for 1.5T scanners being 2.3, 4.6, 6.9, 9.2, 11.5, 13.8 and TE for 3T scanners 1.15, 2.3, 3.45, 4.6, 5.75, 6.9, referred to as \textit{six echo}, resulting in data of 43 MRI scanners being used.

Before randomization and up to 6 weeks before the initial MRI, patients underwent liver biopsy assessing the presence of NASH by the current gold standard of histo-pathological grading of inflammation, hepatocellular ballooning and steatosis. The NAFLD fibrosis score calculated from clinical values (Age, BMI, impaired fasting glucose and/or diabetes, AST, ALT, platelet count and albumin \cite{Angulo2007TheNAFLD}) was used as a fibrosis staging method since fibrosis scoring systems were center specific.
We used RCT data to construct two data sets: \textit{RCT-pred} and \textit{RCT-progress}. An overview of the data set is shown in Table \ref{tab:dataset_overview}.

\begin{table}[ht]
\begin{center}
\begin{small}
\begin{tabular}{llllllll}
                              & Grade        & \textit{\textbf{RCT-pred}} & \textit{\textbf{CR-pred}} &  & \multicolumn{3}{c}{\textit{\textbf{RCT-progress}}}                  \\ \cline{1-4} \cline{6-8} 
\multirow{2}{*}{Diagnosis}    & NASH         & -        & 28      &  & Treat. Group               & Treat. Arm & Patients \\ \cline{6-8} 
                              & Simple Stea. & -        & 18      &  & \multirow{4}{*}{Low Dose}  & 10 mcg     & 7        \\ \cline{1-4}
\multirow{5}{*}{Fibrosis}     & 0            & -        & 10      &  &                            & 30 mcg     & 9        \\
                              & 1            & 44       & 15      &  &                            & 60 mcg     & 22       \\
                              & 2            & 77       & 8       &  &                            & 90 mcg     & 18       \\ \cline{6-8} 
                              & 3            & 11       & 8       &  & \multirow{2}{*}{High Dose} & 140 mcg    & 23       \\
                              & 4            & -        & 5       &  &                            & 200 mcg    & 19       \\ \cline{1-4} \cline{6-8} 
\multirow{4}{*}{Steatosis}    & 0            & 1        & 4       &  & Placebo                    & Placebo    & 47       \\ \cline{6-8} 
                              & 1            & 77       & 20      &  & Total                      &            & 145      \\ \cline{6-8} 
                              & 2            & 37       & 5       &  &                            &            &          \\
                              & 3            & 17       & 17      &  &                            &            &          \\ \cline{1-4}
\multirow{3}{*}{Ballooning}   & 0            & 5        & 23      &  &                            &            &          \\
                              & 1            & 42       & 16      &  &                            &            &          \\
                              & 2            & 85       & 7       &  &                            &            &          \\ \cline{1-4}
\multirow{4}{*}{Inflammation} & 0            & 2        & 17      &  &                            &            &          \\
                              & 1            & 27       & 20      &  &                            &            &          \\
                              & 2            & 31       & 9       &  &                            &            &          \\
                              & 3            & 72       & -       &  &                            &            &          \\ \cline{1-4}
Total                         &              & 132      & 46     &  &                            &            &          \\ \cline{1-4}
\end{tabular}
\end{small}
\end{center}
\caption{\label{tab:dataset_overview}Overview of the data sets used for evaluation and experiments: \textbf{RCT-pred} for prediction of histo-pathology markers and phenotyping, \textbf{RCT-progress} for quantification of progression and treatment response, and \textbf{CR-pred} for replicability analysis on a separate data set.}
\end{table}

\noindent
\textit{\textbf{RCT-pred:}} The data set for biopsy prediction and phenotyping consisted of 132 patients with a complete set of the three MRI sequences (T1w, dixon, six echo), and biopsy-based gradings for steatosis, hepatocyte ballooning and lobular inflammation, as well as the NAFLD fibrosis score. 
\\
\textit{\textbf{RCT-progress:}} To evaluate identification of progression markers, and transition pathways, 145 patients were included in the data set. The patients were randomized into 7 treatment arms with different doses or placebo. Here, we grouped those arms into three treatment groups of low dose (n=56), high dose (n=42) and placebo (n=47) patients. For all patients included in \textit{RCT-progress} three MRI sequences (T1w, dixon and Six Echo) were available at the initial visit and at follow-up after 12 weeks. Note that more patients than in \textit{RCT-pred} could be included in this data set, since we did not require the availability of a biopsy grading.

\subsubsection{Clinical routine replication data set}

We used a separate data set extracted from clinical routine, \textbf{\textit{CR-pred}}, to show the replicability of the approach to different MRI scanners and sequences. \textit{CR-pred} includes not only NASH patients, but overall 46 NAFLD patients, 18 with simple steatosis and 28 with NASH. Biopsy was used to assess the histologic activity for steatosis, inflammation, ballooning and fibrosis (Table~\ref{tab:dataset_overview}).

Imaging data included chemical shift imaging (CSI), with an in-phase and out-of-phase transverse dual echo T1-weighted sequence. Furthermore, native and dynamic contrast-enhanced (with a standard dose of gadoxetic acid), three-dimensional, breath-hold T1-weighted spoiled gradient-echo volumetric (VIBE) sequences were utilized. The contrast-enhanced scan was taken 20 minutes after injection of the contrast agent. All scans were performed on a 3T scanner (Magnetom Trio, A Tim; Siemens Healthcare, Erlangen, Germany). We show results to evaluate the proposed method on CSI images, as well as on a signature-fused combination of native and contrast-enhanced T1w images.

\subsection{Implementation details}
All networks are implemented with Python 3.7.3 and PyTorch 1.4.0 \cite{Paszke2019PyTorch:Library}. For random forest classification and regression the implementation provided by scikit-learn 0.20.3 is used. The code for training and evaluation is available on Github (\href{https://github.com/cirmuw/LiverDCN}{https://github.com/cirmuw/LiverDCN}). 
To obtain the segmentation masks of the liver we employed a U-Net \cite{Ronneberger2015} trained on a subset of manually annotated data. To ensure accuracy and eliminate potential bias from automated segmentation, all masks were subsequently reviewed and corrected by a board-certified radiologist (A.B. or N.B.).
In the following, unless stated otherwise, we use SF-5-3 signatures for all experiments, as this setting has proven to be most universal in preliminary experiments. Other settings are explored in Section \ref{sec:ablation_study}.

\subsection{Identification of non-invasive markers capturing treatment response}
\label{sec:results_progressmarker}
We analysed longitudinal data (\textit{RCT-progress}) and trained a random forest regressor to predict the treatment arm of a patient from the difference between imaging signatures acquired at baseline and after 12 weeks. We evaluated if there is a significant separation between treatment groups, and compared separation with established markers. No parameter tuning of the random forest prediction model was performed. Due to the limited size of the data set, 5-fold cross validation was used to separate training- and test set.

\begin{figure}[t]
\centering
\includegraphics[width=0.95\textwidth]{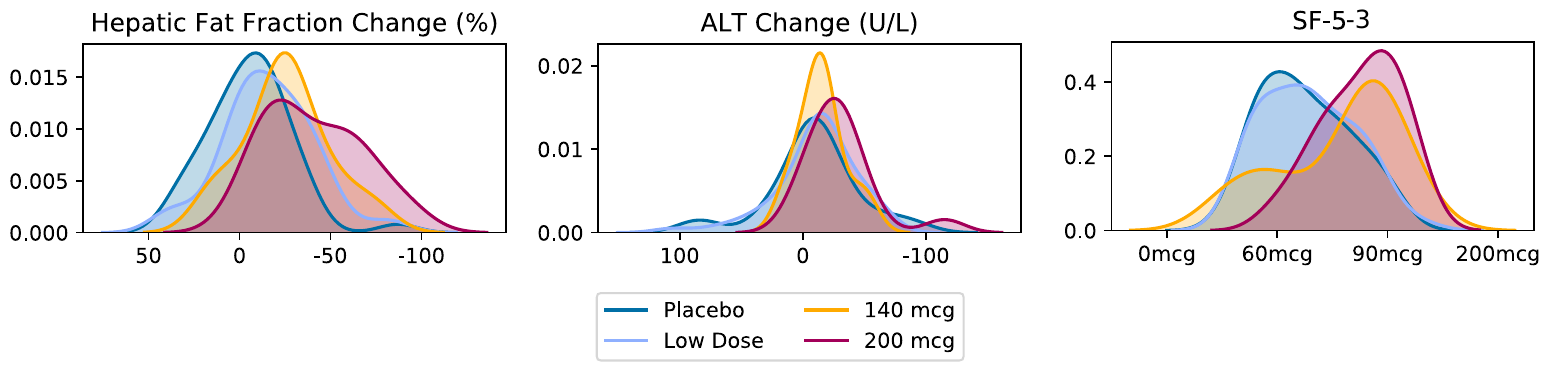}
\caption{Quantitative separation of treatment arms based on MRI signatures compared to established markers. Change of hepatic fat fraction (HFF) and alanine aminotransferase (ALT), two currently used endpoints of the RCT. Note, that the x-axis for change is mirrored HFF and ALT to facilitate comparison. Compared to results for regression of treatment arms from \textit{difference signatures} from SF-5-3 between the initial patient visit and a 12 weeks follow-up. The curves show the predicted density for patients in different treatment arms. Although there is an overlap of the curves, a separation of 140mcg/200mcg and low dose/placebo patients can be observed, as also reflected in the results presented in Table \ref{tab:results_follow_ttest}.}
\label{fig:results_followup}
\end{figure}

Our approach was compared to change of hepatic fat fraction (HFF) and alanine aminotransferase (ALT), which are currently used RCT endpoints.
Figure \ref{fig:results_followup} shows the results for this experiment for our SF-5-3 signatures compared to HFF and ALT. The random forest regressor trained on SF-5-3 resulted in a clear difference between the two groups (placebo and Low dose vs. high doses of 140mcg and 200mcg). This was particularly pronounced for the 200mcg group. Since we expect that not all participants of the trial react to the treatment in the same way, we speculate that part of the overlap between the placebo and high-dose study arms may be attributable to response behavior. However, the comparison with change in hepatic fat fraction  showed that the image signatures were capable of capturing changes specific to the treatment group in more detail than established endpoints.

Table \ref{tab:results_follow_ttest} confirms significant differences between placebo and low dose groups vs. 200mcg dose groups. This indicates that we were able to extract features that could capture disease progression in these groups. We could not find features separating low dose from placebo groups, which is consistent with the low dose treatment used in a phase 2 RCT not leading to an measurable improvement of clinical and laboratory values of the participants \cite{Lucas2020TropifexorResults}. 

\setlength{\tabcolsep}{4pt}
\begin{table}[t]
\centering
\begin{small}
\begin{tabular}{|l|cc|cc|}
\hline
                      \multicolumn{3}{|c|}{Random forest regressor trained on SF-5-3} \\ \hline
                     &  T               & p-corr         \\ \hline
Placebo vs. Low Dose &  0.326         & 1.0      \\ \hline
Placebo vs. 140 mcg  &  2.627         & \textbf{0.06}    \\ \hline
Placebo vs. 200 mcg  &  4.710         & \textbf{0.0001}  \\ \hline
Low Dose vs. 140 mcg &  2.450         & \textbf{0.10}     \\ \hline
Low Dose vs. 200 mcg &  4.448         & \textbf{0.0002}     \\ \hline
140 mcg vs. 200 mcg  &  1.381         & 1.0  \\ \hline
\end{tabular}
\end{small}
\caption{Quantifying response based on signature change. Bonferroni corrected results of t-tests for differences in prediction of treatment arms from longitudinal data. The random forest regressor trained on SF-5-3 can differentiate 140mcg/200mcg against the placebo and low dose groups.}
\label{tab:results_follow_ttest}
\end{table}

\subsection{Identification of liver tissue transition paths under treatment}
\label{sec:results_tissuetransitions}

Tracking the transition of tissue pattern \textit{membership} in longitudinal imaging data enables the identification of paths associated with response. We evaluated changes at each position in the liver from initial patient visits to a 12 week follow-up in the \textit{RCT-progress} data set. We show transitions for SF-5-3 and compare placebo patients to patients treated with high-dose (140 mcg/200 mcg) treatment.

\begin{figure}[ht!]
\centering
\includegraphics[width=0.6\textwidth]{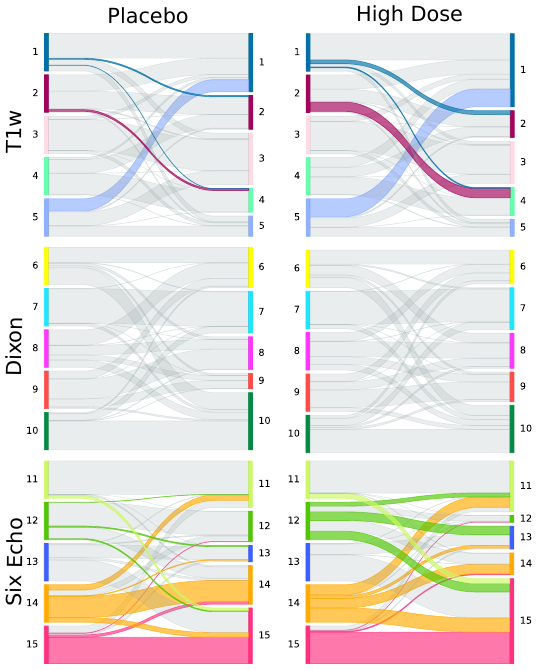}
\caption{Transitions of liver tissue between tissue classes from baseline to follow-up, evaluated with SF-5-3 components. Colors indicate transitions with significantly different frequencies in placebo and high dose patients.}
\label{fig:results_transitions_sf5}
\end{figure}

Figure \ref{fig:results_transitions_sf5} shows the transitions for SF-5-3 tissue patterns in three MR sequences. Differences of transition frequency between placebo and high dose patients are tested with a permutation test and significant differences are marked with colors.
For T1w (cluster 1-5) the pathway from component 2 to 4 was significantly different between the two groups, occurring rarely for placebo subjects, but for 25\% of high dose patients (p=0.02). This was consistent with component 2 being univariately correlated with steatosis (r=0.19, p=0.03), suggesting that the higher dose reduces the amount of steatosis.
The transition from component 5 to component 1 was significantly different (p=0.04). Again this suggests a relationship to component 5 being in the top 5 ranked features for classification of ballooning, inflammation and fibrosis.
For the six echo sequence of SF-5-3 (clusters 11-15) changes in transitions from component 12 and from component 14 are relevant. Both were positively univariately correlated with steatosis (component 12: r=0.19, p=0.03, component 14: r=0.21, p=0.18). While under placebo a majority of 63\% (component 12) and 57\%  (component 14) remained unchanged, for treated patients the transitions were more diverse. There was a significant difference in the transition frequency to component 15 (component 12 to 15: p=0.03, component 14 to 15: p=0.006), which was negatively correlated with steatosis (r=-0.32, p=0.0001). Transitions from tissue positively correlated to tissue negatively correlated with steatosis was more frequent in patients under treatment, consistent with an expected reduction of fat-fraction under treatment.

\subsection{Evaluation of histo-pathology feature prediction from MRI}
\label{sec:results_prediction}

The prediction of histo-pathology features from MRI was evaluated by five-fold cross validation to separate training and test sets on \textit{RCT-pred}. For regression we excluded patients where the stage of the histo-pathology feature is represented in the data set less or equal to five in order to stratify the folds based on the feature. For classification, we report accuracy, positive predictive value (PPV), negative predictive value (NPV), sensitivity and specificity for random forest classification between low and high grade on the \textit{RCT-pred} data set. 
Here, we used the best performing setting of 5 clusters per sequence (SF-5-3) for classification and regression, for different settings see Section \ref{sec:ablation_study}.

In addition, we conducted a replication experiment on \textit{CR-pred}. As for RCT-pred, we extracted \textit{signature-fusion} signatures with 5 clusters per sequence, where the sequences were native and contrast enhanced T1w, as well as a CSI sequence.

\setlength{\tabcolsep}{2pt}
\begin{table}[ht!]
\centering
\begin{small}
\begin{tabular}{lcccccccccc}
\cline{2-11}
\multicolumn{1}{l|}{}          & \multicolumn{5}{c|}{Inflammation}                     & \multicolumn{5}{c|}{Ballooning}                       \\ \cline{2-11} 
\multicolumn{1}{l|}{}          & Acc  & PPV  & NPV  & Sens & \multicolumn{1}{c|}{Spec} & Acc  & PPV  & NPV  & Sens & \multicolumn{1}{c|}{Spec} \\ \hline
\multicolumn{1}{|l|}{RCT-pred} & 0.82 & 0.86 & 0.62 & 0.92 & \multicolumn{1}{c|}{0.45} & 0.73 & 0.78 & 0.63 & 0.81 & \multicolumn{1}{c|}{0.57} \\ \cline{1-1}
\multicolumn{1}{|l|}{CR-pred}  & 0.78 & 0.81 & 0.71 & 0.87 & \multicolumn{1}{c|}{0.63} & 0.70 & 0.74 & 0.67 & 0.61 & \multicolumn{1}{c|}{0.78} \\ \hline
                               &      &      &      &      &                           &      &      &      &      &                           \\ \cline{2-11} 
\multicolumn{1}{l|}{}          & \multicolumn{5}{c|}{Fibrosis}                         & \multicolumn{5}{c|}{Steatosis}                        \\ \cline{2-11} 
\multicolumn{1}{l|}{}          & Acc  & PPV  & NPV  & Sens & \multicolumn{1}{c|}{Spec} & Acc  & PPV  & NPV  & Sens & \multicolumn{1}{c|}{Spec} \\ \hline
\multicolumn{1}{|l|}{RCT-pred} & 0.70 & 0.73 & 0.60 & 0.89 & \multicolumn{1}{c|}{0.34} & 0.74 & 0.76 & 0.73 & 0.54 & \multicolumn{1}{c|}{0.88} \\ \cline{1-1}
\multicolumn{1}{|l|}{CR-pred}  & 0.70 & 0.64 & 0.76 & 0.76 & \multicolumn{1}{c|}{0.64} & 0.85 & 0.86 & 0.84 & 0.82 & \multicolumn{1}{c|}{0.88} \\ \hline
\end{tabular}
\end{small}
\caption{Classification accuracy for low and high grades of biopsy values on both datasets (RCT-pred and CR-pred).}
\label{tab:results_classification}
\end{table}

The evaluation results for prediction of histo-pathology features on both datasets, \textit{RCT-pred} and \textit{CR-pred} are shown in Table~\ref{tab:results_classification}. The results on the two datasets are comparable, which demonstrates that the presented method is robust to the choice of dataset. 
For \textit{CR-pred} in addition to the biopsy-derived markers we can also evaluate the distinction between NASH and Simple Steatosis. Here, our method achieves an accuracy of 0.72 (PPV=0.78, NPV=0.63, Sens=0.75, Spec=0.67). The borders between NASH and Simple Steatosis are challenging to determine for human graders \cite{Vuppalanchi2009EffectsDisease}. This makes the assignment of the ground truth possibly noisy, which could result in reduced accuracy of diagnosis prediction.

\begin{figure}[h]
\centering
\includegraphics[width=0.45\textwidth]{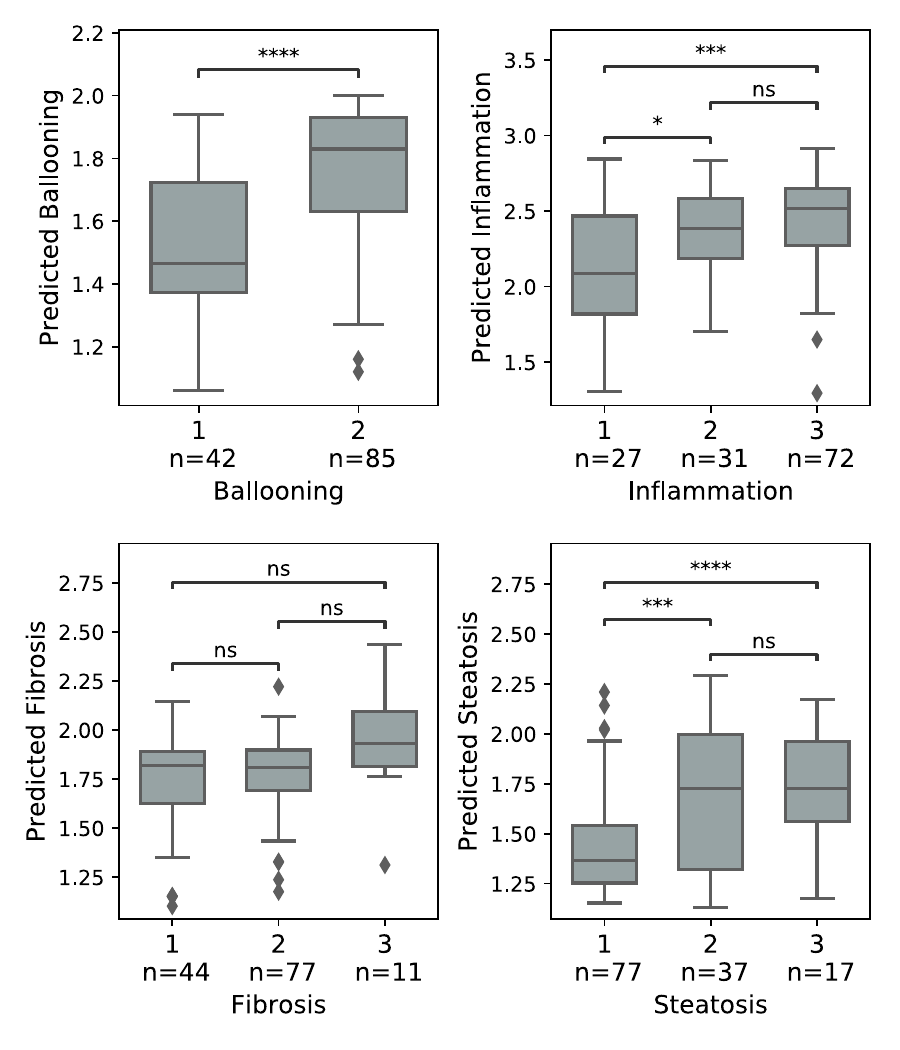} 
\caption{Random forest regression based prediction of biopsy grades of ballooning, inflammation, and steatosis, and NALFD fibrosis score with SF-5-3. T-test with Bonferroni multiple comparison correction was applied: * $0.01<p<=0.05$, ** $0.05<p<=0.001$, *** $0.001<p<=0.0001$, **** $p<0.0001$. Note, that grades represented by five or less patients are excluded, due to the use of five fold cross-validation.}
\label{fig:results_regression}
\end{figure}
Regression results on \textit{RCT-pred} are depicted in Figure \ref{fig:results_regression}. Both fusion strategies showed separation between grade 1 and 2 (p\textless 0.0001) for hepatocyte ballooning. For inflammation, IF-20 struggled with separating grade 1 and 2 while SF-5-3 failed to distinguish between grade 2 and 3. Similarly for steatosis, both fusion strategies failed to separate the higher grades but could successfully make the clinically relevant distinction between low grade and high grade steatosis. On \textit{RCT-pred} fibrosis regression yielded no significant result for predicting the non-invasive fibrosis marker. Although the accuracy of differentiating between biopsy-stages is too low for direct clinical usage, it demonstrates, that the extracted signatures capture relevant characteristics which raises the confidence that the extracted signatures are useful.

\subsection{Association between phenotypes and clinical variables}
\label{sec:results_phenotypes}
\begin{figure}[p]
\centering
\includegraphics[width=\textwidth]{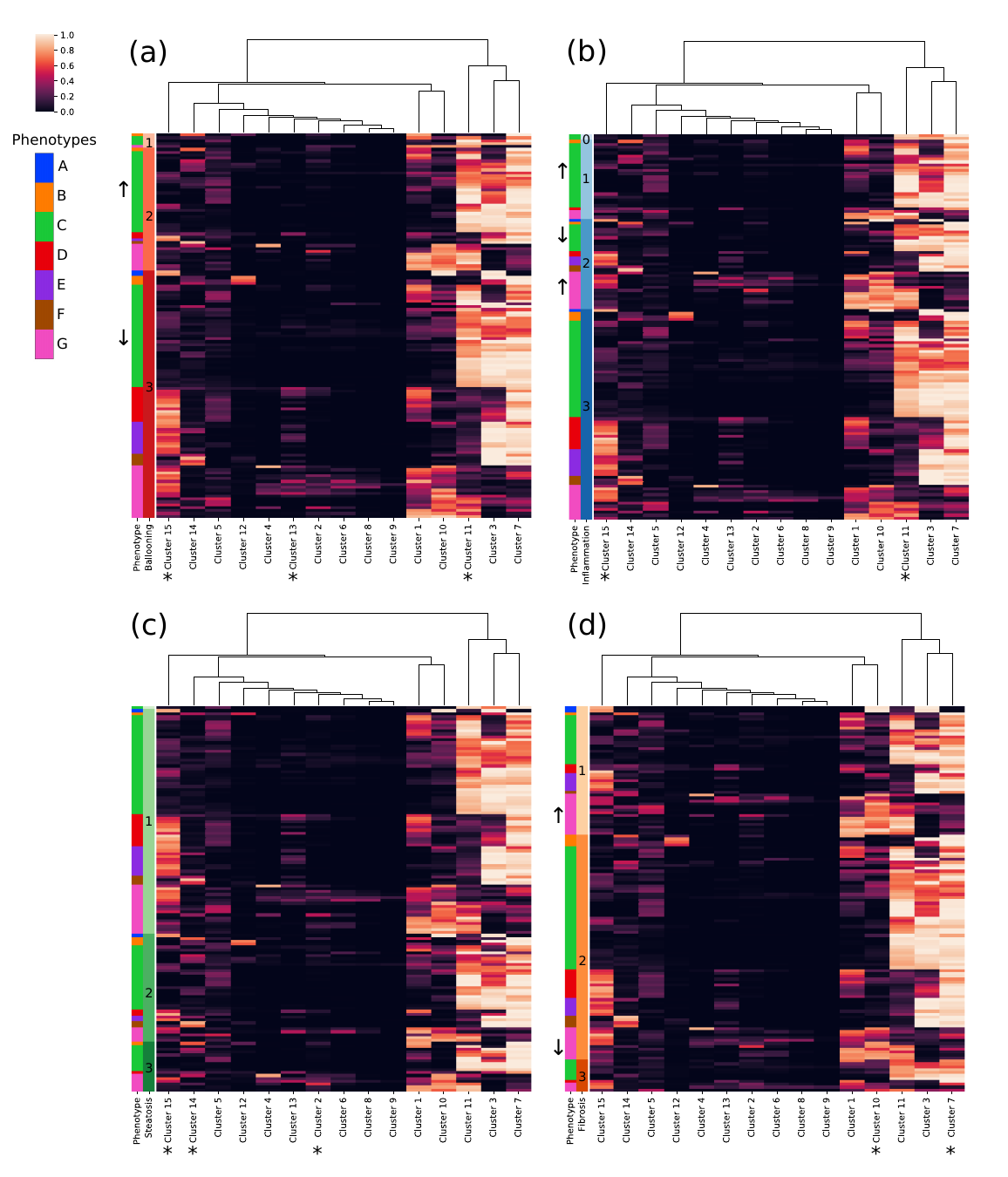}
\caption{Identifying phenotypes and testing their association with histo-pathology markers. Clustermaps of SF-5-3 signatures for image phenotype identification for (a) ballooning, (b) inflammation, (c) steatosis and (d) fibrosis. In each map each row corresponds to one patient, each column one cluster. The coloring of each cell represents the relative frequency of this cluster in a particular patient. The rows are sorted first by biopsy grading and afterwards by a hierarchical clustering over all patients. Hierarchical clustering is also applied to columns to group similar SF-5-3 clusters together. Columns marked with * are significantly (p\textless 0.05) univariate correlated with the corresponding biopsy grade for that feature. $\uparrow$ and $\downarrow$ indicate phenotypes overrepresented and underrepresented, respectively, for a specific clinical grade of a target (see Section \ref{sec:results_phenotypes}).}
\label{fig:sf5_clustermaps}
\end{figure}

To demonstrate the discovery of phenotypes, groups of patients sharing imaging characteristics, associated with clinically relevant parameters, we performed hierarchical, agglomerative clustering with average linkage and euclidean distance on SF-5-3 signatures extracted from \textit{RCT-pred}.
Figure \ref{fig:sf5_clustermaps} depicts maps visualizing the association of these phenotypes with four targets (ballooning, inflammation, steatosis and fibrosis). Thresholding the hierarchical clustering resulted in seven distinct phenotypes (A-G). Phenotype C was characterized by a high amount of components 11, 3 and 7, this phenotype was over represented in patients with ballooning biopsy grade 2 (OR=2.73, p=0.015) and under represented in ballooning biopsy grade 3 (OR=0.36, p=0.01). The same held for inflammation grade 1 (OR=6.1, p\textless 0.0001) and grade 2 (OR=0.32, p=0.013). Phenotype G, characterized by low values for components 3 and 7 and higher values for component 10, was over represented in inflammation grade 2 patients (OR=4.14, p=0.002). For patients with a classified NAFLD fibrosis score of 1 phenotype G was over represented (OR=2.47, p=0.04) and under expressed for score 2 patients (OR=0.37, p=0.03). Although it was possible to discover phenotypes of patients based on the extracted SF-5-3 signatures, no single phenotype was clearly representative of a single disease stage. Instead, composits of phenotypes exhibited an association with composits of disease stages facilitating the interpretation and gaining of insights into the captured features. 

\subsection{Ablation study}
\label{sec:ablation_study}
Here, we analyze the choices of hyperparameter settings and design choice regarding multi-sequence fusion. In Section \ref{sec:fusion_strategy} an alternative to concatenating the signatures from different sequences is explored. Section \ref{sec:cluster_number} discusses the influence of the number of clusters used for signature extraction.

\subsubsection{Ablation of fusion strategy}
\label{sec:fusion_strategy}

\begin{figure}[h!]
\centering
\includegraphics[width=0.6\textwidth]{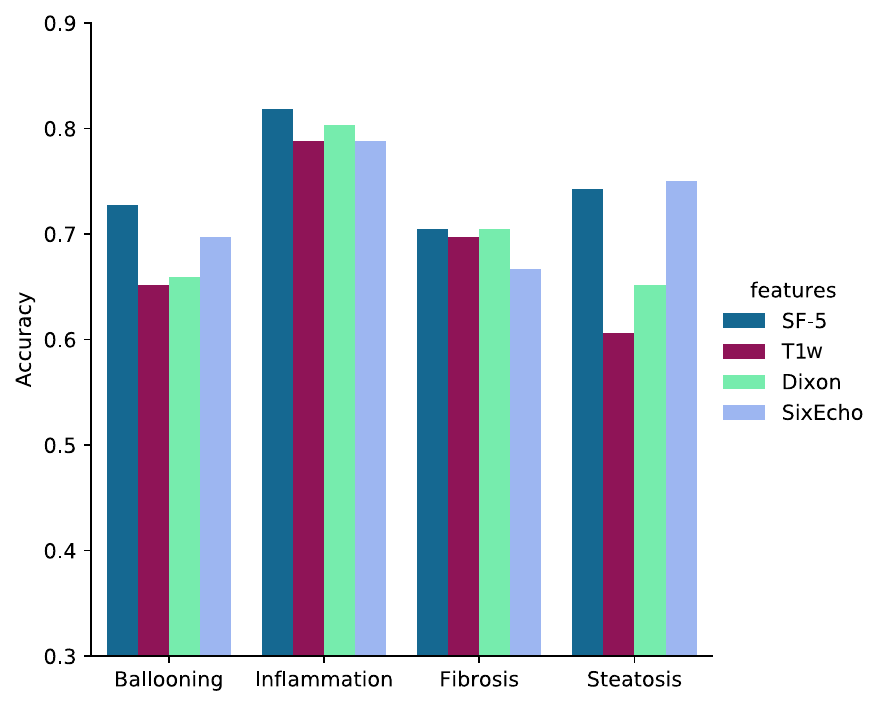}
\caption{Benefit of multi-squence assessment: Comparison of using different sequences on \textit{RCT-pred}. Overall SF-5-3 is the strategy that achieves the best results.}
\label{fig:fused_unfused}
\end{figure}

We test the usage of fusion of the signatures from different sequences on \textit{RCT-pred} by training classifiers on signatures of individual sequences compared to our SF-5-3 signatures. Results are depicted in Figure \ref{fig:fused_unfused}. 

Results demonstrates that fusing multiple sequences is beneficial for all prediction targets.

In addition, as an alternative to signature-fusion we test \textit{\textbf{Image-fusion}}: Images from all sequences are registered to one reference sequence image, so that corresponding image values are available at each position. Patches are then drawn at the same positions across sequences and are concatenated along the channel axis. DCN clustering and \textit{signature} ${\bf s}_{IF}$ calculation is performed once on those multi-sequence patches. We use \textit{IF} as an abbreviation, followed by the cluster number, here we analyse \textit{IF-20} with K=20 image-fused clusters. 

We performed classification and regression results as described in Section \ref{sec:results_prediction}. Results for classification are shown in Table \ref{tab:results_class_image_fusion}, regression results are shown in Figure \ref{fig:results_regression_imagefusion}.

\begin{minipage}{0.45\textwidth}
\vspace{0.8em}
\begin{small}
\begin{tabular}{lcccccc}
\cline{2-7}
\multicolumn{1}{l|}{}          & \multicolumn{3}{c|}{Inflammation}                     & \multicolumn{3}{c|}{Ballooning}                       \\ \cline{2-7} 
\multicolumn{1}{l|}{}          & Acc  & Sens & \multicolumn{1}{c|}{Spec} & Acc  & Sens & \multicolumn{1}{c|}{Spec} \\ \hline
\multicolumn{1}{|l|}{SF-5-3} & 0.82 & 0.92 & \multicolumn{1}{c|}{0.45} & 0.73 & 0.81 & \multicolumn{1}{c|}{0.57} \\ \cline{1-1}
\multicolumn{1}{|l|}{IF-20}  & 0.81 & 0.99 & \multicolumn{1}{c|}{0.17} & 0.76 & 0.92 & \multicolumn{1}{c|}{0.47} \\ \hline
                               &      &      &                           &      &      &                            \\ \cline{2-7} 
\multicolumn{1}{l|}{}          & \multicolumn{3}{c|}{Fibrosis}                         & \multicolumn{3}{c|}{Steatosis}                        \\ \cline{2-7} 
\multicolumn{1}{l|}{}          & Acc   & Sens & \multicolumn{1}{c|}{Spec} & Acc & Sens & \multicolumn{1}{c|}{Spec} \\ \hline
\multicolumn{1}{|l|}{SF-5-3} & 0.70 & 0.89 & \multicolumn{1}{c|}{0.34} & 0.74 & 0.54 & \multicolumn{1}{c|}{0.88} \\ \cline{1-1}
\multicolumn{1}{|l|}{IF-20}  & 0.72 & 0.94 & \multicolumn{1}{c|}{0.27} & 0.64 & 0.48 & \multicolumn{1}{c|}{0.76} \\ \hline
\end{tabular}
\captionof{table}{Classification accuracy for low and high grades of biopsy values for our proposed SF-5-3 signatures compared to the alternative image fusion strategy.}
\label{tab:results_class_image_fusion}
\end{small}
\end{minipage}
\hfill
\begin{minipage}{0.45\textwidth}
\vspace{0.8em}
    \includegraphics[width=\textwidth]{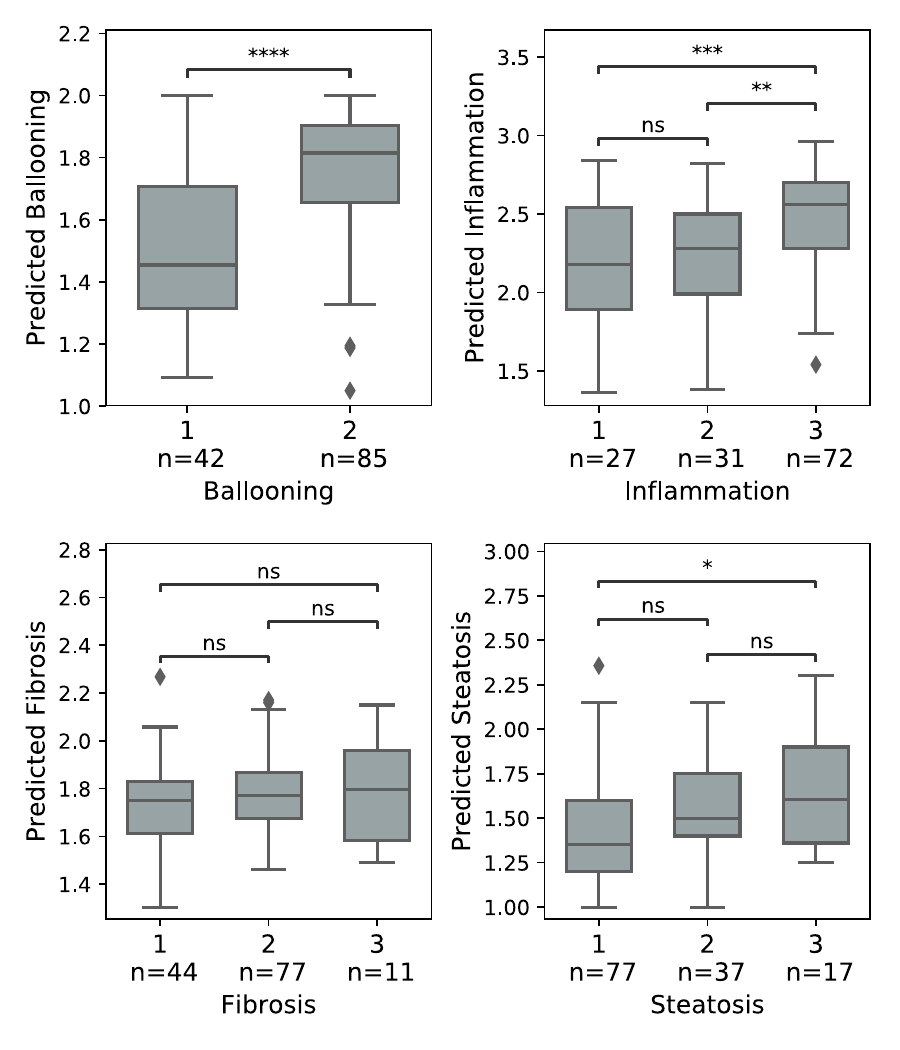}
\captionof{figure}{Random forest regression based prediction of biopsy grades of ballooning, inflammation, and steatosis, and NALFD fibrosis score with SF-5-3. T-test with Bonferroni multiple comparison correction was applied: * $0.01<p<=0.05$, ** $0.05<p<=0.001$, *** $0.001<p<=0.0001$, **** $p<0.0001$. Note, that grades represented by five or less patients are excluded, due to the use of five fold cross-validation.}
\label{fig:results_regression_imagefusion}
\end{minipage}
\vspace{2em}

For classification we observe a similar performance for inflammation, ballooning and fibrosis, for steatosis SF-5-3 is better suitable then IF-20. Those results, in addition to the more simple method of signature-fusion compared to registration when using image-fusion, leads to signature-fusion as the proposed fusion strategy.

\subsubsection{Ablation of cluster numbers}
\label{sec:cluster_number}

The numbers of clusters used is an important hyper-parameter in clustering algorithms, to evaluate the influence of the cluster number for signature-fusion on \textit{RCT-pred}.
We compared the classification performance \textit{signature-fusion} for 5 (our proposed method), 8 and 10 clusters per sequence.
Results are shown in Table \ref{tab:cluster_num_ablation}.

\setlength{\tabcolsep}{2pt}
\begin{table}[ht!]
\centering
\begin{small}
\begin{tabular}{lcccccccccc}
\cline{2-11}
\multicolumn{1}{l|}{}          & \multicolumn{5}{c|}{Inflammation}                     & \multicolumn{5}{c|}{Ballooning}                       \\ \cline{2-11} 
\multicolumn{1}{l|}{}          & Acc  & PPV  & NPV  & Sens & \multicolumn{1}{c|}{Spec} & Acc  & PPV  & NPV  & Sens & \multicolumn{1}{c|}{Spec} \\ \hline
\multicolumn{1}{|l|}{SF-5-3} & 0.82 & 0.86 & 0.62 & 0.92 & \multicolumn{1}{c|}{0.45} & 0.73 & 0.78 & 0.63 & 0.81 & \multicolumn{1}{c|}{0.57} \\ \cline{1-1}
\multicolumn{1}{|l|}{SF-8-3}  & 0.82 & 0.84 & 0.67 & 0.95 & \multicolumn{1}{c|}{0.34} & 0.75 & 0.78 & 0.68 & 0.85 & \multicolumn{1}{c|}{0.57} \\ 
\cline{1-1}
\multicolumn{1}{|l|}{SF-10-3}  & 0.84 & 0.86 & 0.72 & 0.95 & \multicolumn{1}{c|}{0.45} & 0.74 & 0.79 & 0.65 & 0.82 & \multicolumn{1}{c|}{0.60} \\ \hline
                               &      &      &      &      &                           &      &      &      &      &                           \\ \cline{2-11} 
\multicolumn{1}{l|}{}          & \multicolumn{5}{c|}{Fibrosis}                         & \multicolumn{5}{c|}{Steatosis}                        \\ \cline{2-11} 
\multicolumn{1}{l|}{}          & Acc  & PPV  & NPV  & Sens & \multicolumn{1}{c|}{Spec} & Acc  & PPV  & NPV  & Sens & \multicolumn{1}{c|}{Spec} \\ \hline
\multicolumn{1}{|l|}{SF-5-3} & 0.70 & 0.73 & 0.60 & 0.89 & \multicolumn{1}{c|}{0.34} & 0.74 & 0.76 & 0.73 & 0.54 & \multicolumn{1}{c|}{0.88} \\ \cline{1-1}
\multicolumn{1}{|l|}{SF-8-3}  & 0.70 & 0.72 & 0.61 & 0.90 & \multicolumn{1}{c|}{0.32} & 0.74 & 0.74 & 0.74 & 0.57 & \multicolumn{1}{c|}{0.86} \\ 
\cline{1-1}
\multicolumn{1}{|l|}{SF-10-3}  & 0.67 & 0.70 & 0.53 & 0.91 & \multicolumn{1}{c|}{0.20} & 0.67 & 0.68 & 0.67 & 0.39 & \multicolumn{1}{c|}{0.87} \\ \hline
\end{tabular}
\end{small}
\caption{Classification accuracy on \textit{RCT-pred} for low and high grades of biopsy values for 5, 8, and 10 clusters per sequence.}
\label{tab:cluster_num_ablation}
\end{table}

SF showed equal performance for SF-5-3 and SF-8-3, with SF-10-3 suffering a drop of accuracy classifying fibrosis and steatosis. This suggests an overclustering of  tissues, and corresponding noisy feature vectors which weakens the performance of the random forest. None of the settings evaluated outperformed all other settings in every prediction target. Therefore, in our final approach, we propose the most simple setting of using SF-5-3, as a lower number of clusters facilitates interpretability.

\section{Discussion and Conclusion}

In this work we present an unsupervised approach to identify  progression marker patterns in non-invasive medical imaging data. Results suggest that appearance patterns identified in MRI, and tracked during disease progression or treatment can serve as quantitative profiles. They reveal tissue transition paths associated with treatment, and are a step towards non-invasive markers for the development and guidance of therapy.

A deep clustering network trained on medical imaging patches identifies visual tissue patterns occurring frequently in a patient population. These patterns in non-invasive liver MRI of patients suffering from NASH predict markers typically derived from histo-pathology such as steatosis, inflammation, ballooning and fibrosis. A replication experiment suggests robust applicability of the approach. Unsupervised clustering of pattern signatures on a patient level revealed phenotypes associated with disease relevant markers. This indicates that we can identify meaningful groups of patients with similar disease characteristics based on MRI. At the same time, the lack of a one-to-one mapping between marker grades and phenotypes indicates a more complex landscape of patient characteristics captured in multi-parametric MRI.


The utility of these patterns profiles to assess response to treatment was tested by a multivariate regression predicting the treatment dose of a patient from the signature change in longitudinal data. Importantly, this separated treatment groups better than established markers such as ALT or hepatic fat fraction. This suggests that these novel image derived progression and response markers may be a suitable means for quantifying individual progression and response during treatment and clinical trials.

Change can also be tracked locally, to capture the change of tissue properties during progression and treatment. Image registration of follow-up liver MRI data enabled the local tracking of tissue change. This revealed specific tissue classes and corresponding transition behavior associated only with treatment. It allowed for a fine-grained analysis of response, and a focused analysis of underlying mechanisms affecting specific disease tissue types in the liver.


The work has several limitations. The linking of MRI imaging patterns to underlying biology is important, but results have to be interpreted carefully since the gold standard used to train the models (liver biopsy) is prone to sampling variability and intra-reader variability \cite{Ratziu2005SamplingDisease, Vuppalanchi2009EffectsDisease}. This could lead to noisy labels possibly affecting the evaluation. Further analysis how the heterogeneity of tissue properties across the liver is captured by MRI compared to biopsy is necessary. The high diversity of the multi-center clinical trial data set enables a realistic evaluation of the algorithm, but was challenging due to scanner and site variability of the images. Here, we normalize the images, but didn't take any additional measures for normalization across sites into account. Future work will focus on mitigating this limitation.

The present work generates features by unsupervised learning. A direct training and fine-tuning of a classification- or regression model for each of the four tasks might have improved each specific prediction accuracy. However, we evaluated if unsupervised training with a DCN as a task-agnostic model can learn a unified visual tissue pattern representation. It provides marker candidates, associated with different tasks ranging from phenotyping to progression quantification. The ability to predict histo-pathology based markers from these signatures demonstrates that they capture disease related characteristics. It supports the expectation that successful prediction and response quantification are linked to meaningful patterns that can serve as a basis for further research regarding underlying biological mechanisms.


\section*{Acknowledgements}

This work was supported by the Austrian Science Fund (FWF, P 35189), the Vienna Science and Technology Fund (WWTF, LS20-065) and Novartis Pharmaceuticals Corporation. This work has received co-funding from the European Union’s Horizon Europe research and innovation programme under grant agreements No.101136299 — ARTEMIs and No.101080302 AI-POD. M.T. was supported by the Austrian Science Fund (FWF, SFB F73 and the excellence cluster 10.55776/COE14).

\bibliographystyle{IEEEtran}
\bibliography{references}

\end{document}